\documentclass[superscriptaddress,secnumarabic,prd,aps,showpacs,nofootinbib,showkeywords,eqsecnum]{revtex4-2}

\usepackage{graphicx,color,amsmath,amsxtra}
\usepackage{epsf}
\usepackage{amssymb}
\usepackage{enumerate}
\usepackage{hhline}
\usepackage{array}
\usepackage{tabularx}
\usepackage[unicode]{hyperref}
\usepackage{graphicx}                
\usepackage{epstopdf}

\begin{document}
\pagestyle{myheadings}

\title{Anisotropic hyperbolic inflation for a model of two scalar and two vector fields}
\author{Tuan Q. Do }
\email{tuan.doquoc@phenikaa-uni.edu.vn}
\affiliation{Phenikaa Institute for Advanced Study, Phenikaa University, Hanoi 12116, Vietnam}
\affiliation{Faculty of Basic Sciences, Phenikaa University, Hanoi 12116, Vietnam}
\author{W. F. Kao}
\email{homegore09@nycu.edu.tw}
\affiliation{
Institute of Physics, National Yang Ming Chiao Tung University, Hsin Chu 30010, Taiwan
}
\date{\today} 
\begin{abstract}
In this paper, we extend a recent proposed model of two scalar and two vector fields to a hyperbolic inflation scenario, in which the field space of two scalar fields is a hyperbolic space instead of a flat space. In this model, one of the scalar fields is assumed to be a radial field, while the other is set as an angular field. Furthermore, both scalar fields will be coupled to two different vector fields, respectively. As a result, we are able to obtain a set of exact Bianchi type I solutions to this model. Stability analysis is also performed to show that this set of anisotropic solutions is indeed stable and attractive during the inflationary phase. This result indicates that the cosmic no-hair conjecture is extensively violated in this anisotropic hyperbolic inflation model.
\end{abstract}

\maketitle
\section{Introduction} \label{intro}
Cosmic inflation \cite{guth} has been regarded as a leading paradigm in modern cosmology. This result is due to the fact that many of its theoretical predictions have been shown to be highly consistent with the leading cosmic microwave background radiation (CMB) probes such as the Wilkinson Microwave Anisotropy Probe (WMAP) ~\cite{WMAP} and the Planck ~\cite{Planck}. It is worth noting that the backbone of all standard inflationary models \cite{Martin:2013tda} has been the cosmological principle \cite{cosmological-principle}, whose statement is that our universe is just simply homogeneous and isotropic on large scales as described by the Friedmann-Lemaitre-Robertson-Walker (FLRW) spacetime \cite{FLRW}. However, it is not straightforward to verify the validity of this principle \cite{cosmological-principle}. 

It is important to note that some CMB anomalies such as the hemispherical asymmetry and the cold spot have been detected by the WMAP and then confirmed by the Planck \cite{Schwarz:2015cma}. Remarkably, these anomalies are beyond the predictions of all standard inflationary models. It appears that a number of mechanisms, in accordance with the cosmological principle, have been proposed in order to reveal the nature of these anomalies \cite{Schwarz:2015cma}. For instance, there have been some interesting ideas that the CMB statistical anisotropy could be caused by instruments \cite{Hanson:2010gu}. However, a follow-up study has pointed out that they seem to be invalid \cite{Groeneboom:2009cb}. As a result, the physics behind the mentioned CMB anomalies has remained unknown up to now \cite{Schwarz:2015cma}. 
All these results lead us to think of a possibility that the cosmological principle might no longer be valid in the early universe. If so, it might lead to nontrivial deviations from the predictions of standard inflationary models \cite{Pitrou:2008gk}, which might also provide resolutions to other problems. 
For example, it has been shown that the Hubble tension might be an indication of the breakdown of the FLRW cosmology \cite{Krishnan:2021dyb}. 

Remarkably, a recent study has revealed an interesting smoking gun evidence that the current universe might be anisotropic, i.e., might violate the cosmological principle \cite{Colin:2018ghy}. This is indeed contrast to the statement of the so-called cosmic no-hair conjecture proposed by Hawking and his colleagues long ago \cite{Gibbons:1977mu}. The no-hair conjecture states that the late time universe would be homogeneous and isotropic, i.e., would obey the cosmological principle, regardless of initial states of the universe, which might or might not violate the cosmological principle.  
The no-hair conjecture is, however, very difficult to prove. It turns out that there have been a number of partial proofs, e.g., see Refs. \cite{Wald:1983ky,Barrow:1987ia,Kleban:2016sqm,Carroll:2017kjo} for this conjecture since the first rigorous proof by Wald for the Bianchi spacetimes with a cosmological constant, which are homogeneous but  anisotropic \cite{Ellis:1968vb}. Nevertheless, a general proof for this conjecture has remained as a great challenge to physicists  and cosmologists for several decades. It is worth noting that if the cosmic no-hair conjecture is valid, it would only be valid locally, i.e., inside of the future event horizon, according to the studies by Starobinsky and the other people \cite{Starobinsky:1982mr,Barrow:1984zz}. 

Besides the proofs mentioned above, counterexamples to the cosmic no-hair conjecture have been proposed in different frameworks such as higher order models of gravity \cite{Barrow:2005qv}, the Lorentz Chern-Simons model \cite{Kaloper:1991rw}, and the Galileon models \cite{galileon}. However, many of them have been shown to be invalid due to their instability during an inflationary phase \cite{Kao:2009zza}. Recently, the first vivid counterexample to the cosmic no-hair conjecture has been constructed successfully by Kanno, Soda, and Watanabe (KSW) \cite{MW0,MW}. As a result, this counterexample is nothing but a stable and attractive Bianchi type I inflationary solution of a supergravity-motivated model, which involves a special coupling between scalar and vector fields of the form $f^2(\phi)F_{\mu\nu}F^{\mu\nu}$ \cite{MW}. Consequently,  a number of extensions of the KSW model have been proposed in order to either examine the validity of the cosmic no-hair conjecture or investigate the corresponding CMB imprints of anisotropic inflation \cite{extensions,WFK,Do:2017rva,Fujita:2018zbr,multi-vector-1,multi-vector-2,Ohashi:2013pca,non-canonical,Do:2021lyf,Chen:2021nkf,Kim:2013gka,Imprint1,Emami:2015uva}. For interesting reviews on the KSW anisotropic inflation, see Ref. \cite{Maleknejad:2012fw}. It should be noted that the existence of the time-dependent function $f(\phi)$ does break down the conformal invariance of electromagnetic field. Therefore, the KSW anisotropic inflation might have a close connection with the origin of large-scale galactic electromagnetic fields in the present universe as suggested by Refs.  \cite{Turner:1987bw,Ratra:1991bn}. In other words, the appearance of the late time large-scale galactic electromagnetic fields might be a reasonable evidence for the existence of the anisotropic inflationary universe. Moreover, if the KSW anisotropic inflation is stable and viable, then the unavoidable appearance of the late time large-scale galactic electromagnetic fields might be an additional smoking gun evidence for the breaking of the cosmological principle not only in the early universe but also in the late time universe.

Recently, we have proposed a multi scalar and vector fields model, which generalizes many previous extensions of the KSW model \cite{Do:2021lyf}. In this paper, two scalar fields are allowed to non-minimally couple to two vector fields, respectively. Furthermore, this model has been shown to admit an exact Bianchi type I power-law solution, which turns out to be stable and attractive during its inflationary phase. In addition to our model, a recent interesting paper  \cite{Chen:2021nkf} has proposed a different multi-scalar-field extension of the KSW model, which is based on an interesting novel type of inflation called a hyperbolic inflation \cite{Brown:2017osf}. Basically, the hyperbolic inflation model contains two scalar fields, whose two-dimensional field space is hyperbolic instead of a conventional flat one \cite{Easson:2007dh}. One of the scalar fields is referred to as a radial field, while the other one is called an angular field. In this type of inflation, the inflaton, described by the radial field, never slow-rolls and instead orbits the bottom of the potential, buoyed by a centrifugal force \cite{Brown:2017osf}. Consequently, many follow-up works have been done to investigate extensively cosmological aspects of this hyperbolic inflation \cite{Mizuno:2017idt,Bjorkmo:2019aev,Bounakis:2020xaw}. It is noted that only the radial field is non-minimally coupled to a vector field in an anisotropic hyperbolic inflation model proposed in Ref. \cite{Chen:2021nkf}. Naturally, one can ask if the angular field is also non-minimally coupled to a vector field. Apparently, this scenario is similar to our recent model proposed in Ref. \cite{Do:2021lyf}. This motivates us to study in this paper a non-trivial combination of these two extensions of the KSW model. In particular, we will investigate whether an anisotropic hyperbolic inflation \cite{Chen:2021nkf} will appear in a model of two scalar and two vector fields \cite{Do:2021lyf}. Stability analysis will be performed to check if the obtained inflationary solution violates the cosmic no-hair conjecture.

As a result, this paper will be organized as follows: (i) A brief introduction of this study has been presented in Sec. \ref{intro}. (ii) A basic setup of hyperbolic model with two scalar fields coupled to two vector fields will be introduced in Sec. \ref{sec1}. (iii) Anisotropic power-law solutions will be figured out in Sec. \ref{sec2}. (iv) Then, the stability of the obtained solutions will be analyzed using the dynamical system method in Sec. \ref{sec3}. (v) Finally, concluding remarks will be written in Sec. \ref{final}.
\section{The model}  \label{sec1}
In this paper, we would like to study a non-trivial combination of the KSW model \cite{MW0,MW} and the hyperbolic inflation \cite{Brown:2017osf}, which was proposed in Ref. \cite{Chen:2021nkf} such as
\begin{equation} \label{action-0}
S=\int d^4 x\sqrt{-g}  \left[ \frac{1}{2}R -\frac{1}{2} G_{ab}(\phi^a,\phi^b) \partial_\mu \phi^a \partial^\mu\phi^b -V(\phi^a,\phi^b) - \frac{1}{4}f_{ab}(\phi^a,\phi^b) F^a_{\mu\nu} F^{b\mu\nu} \right],
\end{equation}
where the reduced Planck mass $M_p$ has been set to be one for convenience. It is noted that $F^a_{\mu\nu} = \partial_\mu A^a_\nu -\partial_\nu A^a_\mu$ is the field strength of vector field $A^a_\mu$. In addition, $G_{ab}$ is a metric of scalar field space. It should be noted that $f_{ab}$ has been called a gauge kinetic function within the supergravity theory \cite{MW}. However, ones have regarded $f_{ab}$,  in analogy to $G_{ab}$, as a metric of vector field space \cite{Chen:2021nkf}. Both of these metrics have been assumed to be functions of scalar fields in Ref. \cite{Chen:2021nkf}. In this paper, the scalar field and vector field spaces will be assumed to be two-dimensional as
\begin{align}
ds_G^2& =d\phi^2 +\omega L^2 \sinh^2\left(\frac{\phi}{L}\right)  d\psi^2,\\
ds_f^2 &=f_1^2(\phi) d\phi^2 +f_2^2 (\psi) d\psi^2,
\end{align}
respectively.  As a result, the corresponding metrics turn out to be
\begin{align}
G_{ab}& ={\rm diag}\left[1,\omega L^2 \sinh^2 \left(\frac{\phi}{L}\right) \right],\\
f_{ab}&={\rm diag} \left[ f_1^2(\phi),f_2^2 (\psi) \right],
\end{align}
respectively. Here $L>0$ is the curvature scale (length) of the hyperbolic space \cite{Brown:2017osf}, while $\omega =\pm 1$.  Interestingly, the existence of $\omega$ does not affect on the value of the curvature of scalar field space, which is always equal to $-2/L^2 <0$. In other words, the scalar field space is always hyperbolic with negative curvature regardless of the value of $\omega$.

It should be noted that we have renamed $\phi^1 =\phi$ and $\phi^2=\psi$ for convenience. It is noted that $\phi$ is called a radial field, while $\psi$ is called an angular field. It is also noted that $f_1(\phi)$ and $f_2(\psi)$ are arbitrary functions of $\phi$ and $\psi$, respectively. In addition, we will assume in this paper that $V(\phi^1,\phi^2)=V_1(\phi) +V_2(\psi)$, in contrast to Ref. \cite{Chen:2021nkf} where only $V_1(\phi)$ is introduced. It is noted that the configuration of the scalar field space has been proposed in Refs. \cite{Brown:2017osf,Chen:2021nkf}, while the configuration of the vector field space follows our recent paper \cite{Do:2021lyf}, in which two scalar fields are allowed to non-minimally coupled to two vector fields, respectively.  

As a result, the above action \eqref{action-0} now reduces to the following form, 
 \begin{align} \label{action}
S = \int d^4 x\sqrt{-g} \left[ \frac{1}{2}R -\frac{1}{2}\partial_\mu \phi \partial^\mu \phi -\frac{\omega}{2} L^2 \sinh^2\left(\frac{\phi}{L}\right) \partial_\mu \psi \partial^\mu \psi -V_1(\phi) -V_2(\psi) - \frac{f_1^2(\phi)}{4}  F_{\mu\nu}F^{\mu\nu} - \frac{f_2^2(\psi)}{4}  {\cal F}_{\mu\nu} {\cal F}^{\mu\nu} \right] ,
\end{align}
which acts as a hyperbolic generalization of a recent multi-field extension of the KSW model \cite{Do:2021lyf}.  In this action, $F^1_{\mu\nu}\equiv F_{\mu\nu} = \partial_\mu A_\nu -\partial_\nu A_\mu$ is the field strength of the first vector field $A^1_\mu \equiv A_\mu$, while $F^2_{\mu\nu}\equiv{\cal F}_{\mu\nu} = \partial_\mu {\cal A}_\nu -\partial_\nu {\cal A}_\mu$ is the field strength of the second vector field $A^2_\mu\equiv {\cal A}_\mu$. Note that $\psi$ will be a phantom-like scalar field if $\omega$ is equal to $-1$ \cite{Caldwell:1999ew,Guo:2004fq,Chimento:2008ws,Arefeva:2009tkq,Cai:2009zp}.

As a result, varying the action \eqref{action} with respect to the metric $g_{\mu\nu}$ will lead to the corresponding Einstein field equation of this model given by
\begin{align} \label{Einstein-field}
& R_{\mu\nu} -\frac{1}{2}Rg_{\mu\nu } -\partial_\mu \phi \partial_\nu \phi  -\omega L^2 \sinh^2 \left(\frac{\phi}{L}\right) \partial_\mu \psi \partial_\nu \psi \nonumber\\
& +g_{\mu\nu} \left[\frac{1}{2}\partial_\sigma \phi \partial^\sigma \phi +\frac{\omega}{2} L^2 \sinh^2 \left(\frac{\phi}{L}\right) \partial_\sigma \psi \partial^\sigma \psi +V_1+V_2 +\frac{1}{4} \left(f_1^2 F^2+  f_2^2{\cal F}^2 \right) \right] \nonumber\\
&-f_1^2 F_{\mu\gamma}F_\nu{}^{\gamma} -f_2^2 {\cal F}_{\mu\gamma} {\cal F}_\nu{}^{\gamma}=0.
\end{align}
Additionally, the corresponding equations of motion of two vector fields, i.e., $A_\mu$ and ${\cal A}_\mu$, are defined to be
\begin{align} \label{vector-field-1}
\partial_\mu \left[\sqrt{-g}  f_1^2 F^{\mu\nu}  \right] &=0,\\
\label{vector-field-2}
\partial_\mu \left[\sqrt{-g}  f_2^2 {\cal F}^{\mu\nu} \right] &=0,
\end{align}
respectively. 
On the other hand, the corresponding equations of motion of two scalar fields, i.e., $\phi$ and $\psi$, turn out to be
\begin{align} \label{phi-scalar-field}
\square \phi - \frac{\omega}{2} L  \sinh\left(\frac{2\phi}{L}\right) \partial_\sigma \psi \partial^\sigma \psi -\partial_\phi V_1-\frac{1}{2}  f_1 \left(\partial_\phi f_1\right) F^2 &=0,\\
\label{psi-scalar-field}
\omega L^2 \sinh^2 \left(\frac{\phi}{L}\right) \square \psi +\omega L \sinh\left(\frac{2\phi}{L}\right) \partial_\sigma \phi \partial^\sigma \psi -\partial_\psi V_2-\frac{1}{2}f_2 \left(\partial_\psi f_2\right) {\cal F}^2 &=0,
\end{align}
respectively. It is noted that $\partial_\phi \equiv \partial/\partial \phi $, $\partial_\psi \equiv \partial/\partial \psi $, and $\square \equiv \frac{1}{\sqrt{-g}} \partial_\mu \left(\sqrt{-g} \partial^\mu \right)$. 
In this paper, we would like  to figure out anisotropic hyperbolic solutions to this model. To do this task, we will consider the Bianchi type I metric, which is considered as the simplest homogeneous but anisotropic spacetime having the following form \cite{MW0,MW}
\begin{equation} \label{metric}
ds^2 =-dt^2 +\exp\left[ 2\alpha(t) -4\sigma(t) \right] dx^2 +\exp\left[ 2\alpha(t) +2\sigma(t) \right] \left(dy^2+dz^2 \right),
\end{equation}
where $\sigma(t)$ is assumed to be a deviation from the spatial isotropy, which is governed by $\alpha(t)$. This assumption corresponds to a sufficient condition  that $\sigma(t)$ should be much smaller than  $\alpha(t)$ during an inflationary phase.  In accordance with the Bianchi type I metric having the $y-z$ rotational symmetry as shown in Eq. \eqref{metric}, the configuration of two vector fields, $A_\mu$  and ${\cal A}_\mu$, will be considered as  $A_\mu   = \left( {0,A_x \left( t \right),0,0} \right)$ and  ${\cal A}_\mu   = \left( {0,{\cal A}_x \left( t \right),0,0} \right)$. Additionally, both scalar fields will be regarded as homogeneous ones, i.e., they will only be functions of cosmic time, $\phi =\phi(t)$ and $\psi=\psi(t)$.

As a result, the corresponding solutions of vector field equations, i.e., Eqs. \eqref{vector-field-1} and \eqref{vector-field-2}, turn out to be
\begin{align}
\dot A_x & =p_A f_1^{-2}  \exp[-\alpha-4\sigma],\\
{\dot {\cal A}}_x &=q_A f_2^{-2} \exp[-\alpha-4\sigma],
\end{align}
respectively. Here, $p_A$ and $q_A$ are integration constants.  Thanks to these solutions, the field equations \eqref{Einstein-field}, \eqref{phi-scalar-field}, and \eqref{psi-scalar-field} can be rewritten explicitly as follows
\begin{align} \label{field-equation-1}
\dot\alpha^2 &= \dot\sigma^2 +\frac{1}{3} \left[ \frac{\dot\phi^2}{2}+\frac{\omega}{2} L^2 \sinh^2 \left(\frac{\phi}{L}\right) \dot\psi^2+V_1 +V_2 +\frac{1}{2} \left(p_A^2 f_1^{-2}+q_A^2 f_2^{-2}  \right)\exp[-4\alpha-4\sigma] \right],\\
\label{field-equation-2}
\ddot\alpha&=-3\dot\alpha^2 +V_1+V_2 +\frac{1}{6} \left(p_A^2 f_1^{-2}+q_A^2 f_2^{-2}  \right) \exp[-4\alpha-4\sigma],\\
\label{field-equation-3}
\ddot\sigma&=-3\dot\alpha \dot\sigma +\frac{1}{3} \left(p_A^2 f_1^{-2}+q_A^2 f_2^{-2}  \right) \exp[-4\alpha-4\sigma],\\
\label{field-equation-4}
\ddot\phi&=-3\dot\alpha \dot\phi + \frac{\omega}{2} L  \sinh\left(\frac{2\phi}{L}\right) \dot\psi^2 -\partial_\phi V_1 +  p_A^2 f_1^{-3}(\partial_\phi f_1)\exp[-4\alpha-4\sigma],\\
\label{field-equation-5}
\ddot\psi&=-3\dot\alpha \dot\psi -\frac{2}{L}\coth \left(\frac{\phi}{L}\right)\dot\phi \dot\psi  - \frac{1}{\omega L^2 \sinh^2 \left(\frac{\phi}{L}\right)} \left[ \partial_\psi V_2 - q_A^2 f_2^{-3} (\partial_\psi f_2) \exp[-4\alpha-4\sigma] \right] .
\end{align} 
It turns out that we now have five equations for four variables, $\alpha$, $\sigma$, $\phi$, and $\psi$. However, it should be noted that Eq. \eqref{field-equation-1} is nothing but the Friedmann equation, which just plays as a constraint field equation. On the other hand, the time evolution of the spatial isotropy $\alpha$ will be described by Eqs. \eqref{field-equation-2}, while that of the spatial anisotropy $\sigma$  will be determined by Eq. \eqref{field-equation-3}.
\section{Power-law solutions for anisotropic hyperbolic inflation} \label{sec2}
It turns out that the above field equations are difficult to be solved to give a power-law inflation \cite{Abbott:1984fp} due to the existence of hyperbolic functions such as $\sinh(\phi/L)$ and $\coth(\phi/L)$. However, as suggested in Ref.  \cite{Chen:2021nkf} it is possible to figure out power-law solutions in the regime $\phi\gg L$. It is due to the result that the hyperbolic functions can be approximated as exponential functions in this regime,
\begin{equation}
\sinh \left(\frac{\phi}{L}\right) \simeq \cosh \left(\frac{\phi}{L}\right) \simeq \frac{1}{2} \exp\left(\frac{\phi}{L}\right); ~\sinh \left(\frac{2\phi}{L}\right) \simeq  \frac{1}{2}\exp\left(\frac{2\phi}{L}\right);~ \coth \left(\frac{\phi}{L}\right)\simeq 1.
\end{equation}
In this paper, we would like to figure out power-law solutions by choosing the following ansatz \cite{MW,WFK,Chen:2021nkf,Do:2021lyf}
\begin{equation}
\alpha (t)= \zeta \log t; ~\sigma(t) = \eta \log t;~ \phi(t) = \xi \log t +\phi_0;~\psi(t) = \psi_0 t^p,
\end{equation}
together with the compatible potential and coupling functions, whose forms are given by \cite{Chen:2021nkf,Do:2021lyf}
\begin{align}
V_1(\phi)&=V_{01} \exp[\lambda \phi],\\
V_2(\psi)&=V_{02} \psi^n,\\
f_1(\phi)&=f_{01}\exp[\rho \phi],\\
f_2(\psi)&=f_{02}\psi^m,
\end{align}
here $\phi_0$, $\psi_0$, $\xi_i$, $V_{0i}$, $f_{0i}$, $\lambda$, $\rho$, $n$, and $m$ are all non-vanishing parameters. As a result, the scale factors are now of power-law functions as
\begin{equation}
\exp[2\alpha-4\sigma]=t^{2\zeta-4\eta}; ~\exp[2\alpha+2\sigma] =t^{2\zeta+2\eta}.
\end{equation}
As a result, the value of $\zeta$ and $\eta$ will tell us how fast the expansion of our universe is. In particular, it appears that $\zeta-2\eta>0$ and $\zeta+\eta>0$ are two sufficient constraints for expanding universe, while $\zeta-2\eta \gg 1$ and $\zeta+\eta\gg 1$ are two sufficient constraints for inflationary universe \cite{MW,WFK}.

As a result, a set of algebraic equations is defined, in  the regime $\phi \gg L$, from the above field equations  to be
\begin{align}
\label{algebraic-1}
\zeta^2 &=\eta^2 +\frac{1}{3} \left[ \frac{\xi^2}{2} +\frac{\omega}{2} u_0 p^2    +u_1 +u_2 +\frac{1}{2} \left(v_1 +v_2 \right) \right],\\
\label{algebraic-2}
-\zeta &= -3\zeta^2+ u_1 +u_2 +\frac{1}{6} \left(v_1 +v_2 \right),\\
\label{algebraic-3}
-\eta&=-3\zeta\eta +\frac{1}{3} \left(v_1 +v_2 \right),\\
\label{algebraic-4}
-\xi &=-3\zeta \xi +\frac{\omega}{L} u_0 p^2   -\lambda u_1 +\rho v_1,\\
\label{algebraic-5}
 p \left(p-1\right)&= -3  \zeta p   -\frac{2}{L}  \xi p  -\frac{1}{\omega   u_0 }\left(n u_2-mv_2\right),
\end{align}
where additional variables $u_i$ and $v_i$ have been introduced as
\begin{align} \label{definition-of-u0}
u_0&= \frac{1}{4}L^2\psi_0^2 \exp \left(\frac{2\phi_0}{L}\right),\\
u_1&= V_{01}\exp[\lambda \phi_0 ],\\
u_2&= V_{02} \psi_0^n,\\
v_1&=p_A^2 f_{01}^{-2}\exp[-2\rho \phi_0],\\
v_2&=q_A^2 f_{02}^{-2}\psi_0^{-2m}.
\end{align}
It is noted that the following constraints,
\begin{align}
\label{constraint-1}
\frac{ \xi}{L}+p&=0,\\
\label{constraint-2}
\lambda \xi& =-2,\\
\label{constraint-3}
np&=-2,\\
\label{constraint-4}
\zeta+\eta + \frac{1}{2}\rho \xi & =\frac{1}{2},\\
\label{constraint-5}
\zeta+\eta +\frac{1}{2}mp&=\frac{1}{2},
\end{align}
have been used to define the above set of algebraic equations. As a result, the last two constraints shown in Eqs. \eqref{constraint-4} and \eqref{constraint-5} imply that an useful relation
\begin{equation} \label{relation-1}
\rho \xi =mp.
\end{equation}
Furthermore, this relation can be simplified to 
\begin{equation} \label{relation-2-1}
\frac{\rho}{\lambda} =\frac{m}{n} =\kappa_1,
\end{equation}
with the help of the other constraints shown in Eqs. \eqref{constraint-2} and \eqref{constraint-3}. Consequently, it appears that
\begin{equation} \label{relation-2-2}
n\rho =m\lambda =\kappa_2.
\end{equation}
Here $\kappa_1$ and $\kappa_2$ are additional constants. 
On the other hand, both constraint equations \eqref{constraint-4} and \eqref{constraint-5} imply that 
\begin{equation} \label{relation-3}
\zeta = \kappa_1 -\eta +\frac{1}{2}.
\end{equation}
According to our recent paper \cite{Do:2021lyf}, we introduce two additional variables
\begin{align}
&u=u_1+u_2,\\
 &v=v_1+v_2,
\end{align}
for convenience. As a result, $u$ and $v$ can be figured out from two equations, \eqref{algebraic-2} and \eqref{algebraic-3}, as
\begin{align}
\label{definition-of-u}
&u=\zeta \left(3\zeta-1\right)-\frac{v}{6},\\
\label{definition-of-v}
&v= 3\eta \left(3\zeta-1\right),
\end{align}
respectively. 
As a result, we can further simplify Eq. \eqref{algebraic-5} as
\begin{equation} \label{algebraic-6}
 - p = -3  \zeta p  +p^2 -\frac{1}{\omega   u_0 }\left(n u_2-mv_2\right),
 \end{equation}
 with the help of the constraint \eqref{constraint-1}.  
 Furthermore, combining this equation with Eq. \eqref{algebraic-4} will lead to 
 \begin{equation} \label{equation-of-zeta-1}
 \left(3\zeta-1\right) \left(m\xi +\omega  u_0 \rho p  \right) = -\kappa_2 u +m\rho v,
 \end{equation}
 with the help of the relation $\rho L =-m$ derived from Eqs. \eqref{constraint-1} and \eqref{relation-1}. As a result, plugging the $u$ and $v$ defined in Eqs. \eqref{definition-of-u} and \eqref{definition-of-v} into Eq. \eqref{equation-of-zeta-1} leads to the corresponding equation of $\zeta$,
 \begin{equation}
 \left(3\zeta-1 \right) \left[ 6n \lambda \left(\kappa_2 +2m\rho \right) \zeta - n\lambda \left(2\kappa_1+1 \right) \left(\kappa_2 +6 m\rho \right)-8 \left(\omega u_0 \lambda \rho +mn \right) \right]=0,
 \end{equation}
 which can be solved to give a non-trivial solution of $\zeta$,
 \begin{equation} \label{solution of zeta}
 \zeta = \frac{n\lambda \left(2\kappa_1+1 \right) \left(\kappa_2 +6 m\rho \right)+8 \left(\omega u_0 \lambda \rho +mn \right)}{6n \lambda \left(\kappa_2 +2m\rho \right) }.
 \end{equation}
 As a result, this solution does satisfy Eq. \eqref{algebraic-1} derived from the Friedmann constraint equation, regardless of non-vanishing value of $u_0$.  It turns out that this solution is similar to that found in a non-hyperbolic inflation model \cite{Do:2021lyf}, in which $V_2(\psi)=V_{02} \exp[\lambda_2 \psi]$, $f_2(\psi) =f_{02}\exp[\rho_2 \psi] $, and $\psi =\xi_2 \log t+\psi_0$. To be more specific, there is a correspondence that $n\sim \lambda_2$ and $m \sim \rho_2$ between the two solutions of two different models, one is hyperbolic and the other is non-hyperbolic \cite{Do:2021lyf}. 
 
 Given the solution of $\zeta$, the corresponding $\eta$ is defined to be
 \begin{equation} \label{solution of eta}
 \eta =  \frac{\kappa_2 n \lambda \left(2\kappa_1+1\right) - 4 \left(\omega u_0 \lambda \rho +mn \right) }{3n\lambda \left(\kappa_2 +2m \rho \right)}.
 \end{equation}
 For the value of $L$, it appears from Eq. \eqref{constraint-1} that
 \begin{equation} \label{L-relation}
 L= -\frac{n}{\lambda}=-\frac{m}{\rho},
 \end{equation}
 with the help of Eqs. \eqref{relation-1} and \eqref{relation-2-1}. As a result, the positivity of $L$ implies that all $n$ and $m$ should be negative definite since $\rho$ and $\lambda$ are both assumed to be positive definite. It appears that if $|m| \sim \rho $ as well as $|n|\sim \lambda $ then $L \sim {\cal O}(1)$. On the other hand, if $\rho \gg |m| $ as well as $\lambda \gg |n| $ then $L \ll 1$.
  
Now, we would like to see whether these solutions represent inflationary one. As a result, the inflationary constraints, $\zeta+\eta\gg 1$ and $\zeta-2\eta \gg 1$ can be easily fulfilled if $\rho \gg \lambda$ along with $|m| \gg |n|$. Consequently, we have the following approximations as
 \begin{align}
 &\zeta \simeq \kappa_1 \gg 1,\\
 &\eta \simeq \frac{1}{3} ,\\
 & u\simeq 3\kappa_1^2,\\
 &v\simeq 3\kappa_1.
 \end{align}
 In conclusion, an exact power-law solution of anisotropic hyperbolic inflation having a small spatial anisotropy, $\Sigma/H \equiv \dot\sigma/\dot\alpha = \eta/\zeta \simeq 1/(3\kappa_1) \ll 1$, has been figured out in the regime that $\phi \gg L$. More interestingly, this solution turns out to be similar to that found in the recent non-hyperbolic two scalar and two vector fields model \cite{Do:2021lyf}. Now, we would like to compare the present inflationary solution with the solutions found in Ref. \cite{Chen:2021nkf}. It appears that, when $V_2(\psi)$ and $f_2(\psi)$ are removed altogether, the corresponding anisotropic hyperbolic inflation for one scalar-vector coupling has been given by \cite{Chen:2021nkf}
 \begin{align}
 \zeta_0 = \frac{1}{3} \left(\frac{2}{L\lambda}+1 \right),\\
 \eta_0 =\frac{1}{6} +\frac{\rho}{\lambda} -\frac{2}{3L\lambda},
 \end{align}
 where $L$ now acts as a free parameter. Therefore, it is clear that $\zeta_0 \simeq \rho/\lambda \simeq \zeta \gg 1 $ as well as $\eta_0 \sim 1/6$ during an inflationary phase, provided that $\rho/\lambda \sim 2/(3L\lambda)$. This result implies that the existence of the potential $V_2(\psi)$ and the additional coupling between the angular and second vector fields, i.e., $f_2^2(\psi){\cal F}^2$, does not modify significantly the value of scale factors of the metric.
  In the next section, we will see whether this solution is stable or not. Additionally, we will numerically examine whether it is attractive or not. This is an important task in order to check the validity of the cosmic no-hair conjecture. 
 \section{Stability analysis} \label{sec3}
 In this section, we would like to investigate the stability of the obtained anisotropic power-law hyperbolic inflationary solution. It should be noted that in the present model both $V_2(\psi)$ and $f_2(\psi)$ have been assumed as power-law functions of $\psi$. Hence, we should define the corresponding suitable dynamical variables, which might not be introduced in the previous paper, where both $V_2(\psi)$ and $f_2(\psi)$ are exponential functions of $\psi$  \cite{Do:2021lyf}. Fortunately, this issue can be easily handled thanks to  some earlier works  investigating dynamical systems for cosmological models having power-law potentials of scalar field \cite{Bahamonde:2017ize}. As a result, we will define, hinted by Refs. \cite{MW,Do:2021lyf,Chen:2021nkf,Guo:2004fq,Bahamonde:2017ize}, the corresponding dimensionless dynamical variables  as follows 
 \begin{align}
& X=\frac{\dot\sigma}{\dot\alpha}; ~Y_1 = \frac{\dot\phi}{\dot\alpha}; ~Y_2 = \frac{L}{2}\exp\left(\frac{\phi}{L} \right)\frac{\dot\psi}{\dot\alpha},\\
&Z_1=\frac{p_A f_1^{-1}}{\dot\alpha} \exp[-2\alpha-2\sigma],\\
 &Z_2=\frac{q_A f_2^{-1}}{\dot\alpha} \exp[-2\alpha-2\sigma],\\
&W_1=\frac{\sqrt{V_1}}{\dot\alpha};~W_2=\frac{\sqrt{V_2}}{\dot\alpha}, \\
&U_1 = \frac{\bar\lambda}{\bar\lambda+1}; ~U_2 =\frac{\bar\rho}{\bar\rho+1},
 \end{align}
 where $\bar\lambda$ and $\bar\rho$ are defined as
 \begin{equation}
 \bar\lambda = \frac{2}{L} \exp\left(-\frac{\phi}{L} \right) \frac{\partial_\psi V_2}{V_2}; ~\bar\rho =  \frac{2}{L} \exp\left(-\frac{\phi}{L} \right) \frac{\partial_\psi f_2}{f_2}.
 \end{equation}
Here, $W_1$, $W_2$, $U_1$, and $U_2$ are auxiliary dynamical variables, which help us to have a complete dynamical system \cite{Guo:2004fq,Bahamonde:2017ize}. It is noted that the definition of $\bar\lambda$ and $\bar\rho$ for non-hyperbolic models should not involve $2L^{-1}\exp\left(-{\phi}/{L} \right)$ \cite{Bahamonde:2017ize}. It is clear that if both $V_2(\psi)$ and $f_2(\psi)$ are exponential functions of $\psi$ as proposed in a non-hyperbolic inflation model \cite{Do:2021lyf} then both $\bar\lambda$ and $\bar\rho$ will be constant. Consequently, both $U_1$ and $U_2$ will also be constant and therefore cannot be dynamical variables. That is a reason why we did not introduce them in the previous paper \cite{Do:2021lyf}.

As a result, we are able to have the following autonomous equations for the present model,
 \begin{align}
\frac{dX}{d\alpha} &= \frac{\ddot\sigma}{\dot\alpha^2} -\frac{\ddot\alpha}{\dot\alpha^2}X,\\
\frac{dY_1}{d\alpha}&=\frac{\ddot\phi}{\dot\alpha^2}-\frac{\ddot\alpha}{\dot\alpha^2}Y_1,\\
\frac{dY_2}{d\alpha}&= \frac{L}{2}\exp\left(\frac{\phi}{L} \right) \frac{\ddot\psi}{\dot\alpha^2} +\left(\frac{Y_1}{L}- \frac{\ddot\alpha}{\dot\alpha^2}\right) Y_2,\\
\frac{dZ_1}{d\alpha}&= - \left[2\left(X+1\right)+ \rho Y_1 + \frac{\ddot\alpha}{\dot\alpha^2}\right] Z_1,\\
\frac{dZ_2}{d\alpha}&= - \left[2\left(X+1\right)+ \frac{U_2}{1-U_2}Y_2 +\frac{\ddot\alpha}{\dot\alpha^2} \right] Z_2 ,\\
\frac{dW_1}{d\alpha}&= \left(\frac{\lambda}{2}Y_1 -\frac{\ddot\alpha}{\dot\alpha^2} \right)W_1,\\
\frac{dW_2}{d\alpha}&=  \left(\frac{U_1}{1-U_1} \frac{Y_2}{2} - \frac{\ddot\alpha}{\dot\alpha^2} \right)W_2,\\
\frac{dU_1}{d\alpha}&=- \left( \frac{1-U_1}{U_1}\frac{Y_1}{L} + \frac{Y_2}{n} \right)U_1^2 , \\
\frac{dU_2}{d\alpha}&=-  \left( \frac{1-U_2}{U_2}\frac{Y_1}{L} + \frac{Y_2}{m}  \right) U_2^2 ,
\end{align}
where $\alpha$ plays as a new time coordinate related to the cosmic time $t$ as $d\alpha=\dot\alpha dt$.  As a result, using the field equations obtained in the previous section, i.e., Eqs. \eqref{field-equation-2}, \eqref{field-equation-3}, \eqref{field-equation-4}, and \eqref{field-equation-5}, we will write down the explicit autonomous equations of dynamical system as follows 
\begin{align}
\label{dynamical-1}
\frac{dX}{d\alpha} &= X \left[3\left(X^2-1 \right) +\frac{1}{2} \left(Y_1^2 +\omega Y_2^2 \right)+\frac{1}{3} \left(Z_1^2 +Z_2^2 \right) \right] +\frac{1}{3} \left(Z_1^2+Z_2^2 \right),\\
\label{dynamical-2}
\frac{dY_1}{d\alpha}&=Y_1 \left[3\left(X^2-1 \right) +\frac{1}{2} \left(Y_1^2 +\omega Y_2^2 \right)+\frac{1}{3} \left(Z_1^2 +Z_2^2 \right) \right] +\frac{\omega}{L}Y_2^2  +\rho Z_1^2 -\lambda W_1^2,\\
\label{dynamical-3}
\frac{dY_2}{d\alpha}&= Y_2 \left[3\left(X^2-1 \right) +\frac{1}{2} \left(Y_1^2 +\omega Y_2^2 \right)+\frac{1}{3} \left(Z_1^2 +Z_2^2 \right) \right] -\frac{1}{L}Y_1Y_2  +\frac{1}{\omega} \frac{U_2}{1-U_2} Z_2^2 -\frac{1}{\omega} \frac{U_1}{1-U_1} W_2^2,\\
\label{dynamical-4}
\frac{dZ_1}{d\alpha}&= Z_1 \left[ 3\left(X^2-1 \right) +\frac{1}{2} \left(Y_1^2 +\omega Y_2^2 \right)+\frac{1}{3} \left(Z_1^2 +Z_2^2 \right)  -2X -\rho Y_1 +1 \right],\\
\label{dynamical-5}
\frac{dZ_2}{d\alpha}&= Z_2 \left[ 3\left(X^2-1 \right) +\frac{1}{2} \left(Y_1^2 +\omega Y_2^2 \right)+\frac{1}{3} \left(Z_1^2 +Z_2^2 \right)  -2X -\frac{U_2}{1-U_2} Y_2 +1 \right],\\
\label{dynamical-6}
\frac{dW_1}{d\alpha}&= W_1 \left[ 3X^2 +\frac{1}{2} \left(Y_1^2 +\omega Y_2^2 \right)+\frac{1}{3} \left(Z_1^2 +Z_2^2 \right)  +\frac{\lambda}{2}Y_1 \right],\\
\label{dynamical-7}
\frac{dW_2}{d\alpha}&=  W_2 \left[ 3X^2 +\frac{1}{2} \left(Y_1^2 +\omega Y_2^2 \right)+\frac{1}{3} \left(Z_1^2 +Z_2^2 \right)  +\frac{U_1}{1-U_1}\frac{Y_2}{2} \right],\\
\label{dynamical-8}
\frac{dU_1}{d\alpha}&=- \left( \frac{1-U_1}{U_1}\frac{Y_1}{L} + \frac{Y_2}{n} \right) U_1^2 , \\
\label{dynamical-9}
\frac{dU_2}{d\alpha}&=-  \left( \frac{1-U_2}{U_2}\frac{Y_1}{L} + \frac{Y_2}{m}  \right)U_2^2  .
\end{align}
It is noted that the useful relation, 
\begin{equation}\label{dynamical-constraint}
W_1^2+W_2^2 =-3\left(X^2-1 \right)-\frac{1}{2}\left(Y_1^2+\omega Y_2^2 \right)-\frac{1}{2} \left(Z_1^2+Z_2^2 \right),
\end{equation}
which is obtained from the Friedmann equation \eqref{field-equation-1}, has been used to derive the above dynamical system.  Now, we would like to seek anisotropic fixed points with $X\neq 0$ to this dynamical system and study their attractive property. Mathematically, fixed points of the dynamical system, which can be isotropic or anisotropic, are solutions of the following set of equations, 
\begin{equation} \label{dynamical-10}
\frac{dX}{d\alpha}=\frac{dY_1}{d\alpha}=\frac{dY_2}{d\alpha} =\frac{dZ_1}{d\alpha}=\frac{dZ_2}{d\alpha}=\frac{dW_1}{d\alpha}=\frac{dW_2}{d\alpha}=\frac{dU_1}{d\alpha}=\frac{dU_2}{d\alpha}=0.
\end{equation}
As a result, two equations, ${dU_1}/{d\alpha}={dU_2}/{d\alpha}=0$, give  us a relation
\begin{equation} \label{dynamical-11}
\frac{m}{n} =\frac{U_2 \left(U_1-1\right)}{U_1 \left(U_2-1\right)},
\end{equation}
provided a requirement that $U_1 \neq 0$ and $U_2 \neq 0$. Furthermore, this relation can be reduced to a relation between $U_1$ and $U_2$ as
\begin{equation} \label{dynamical-12}
U_2 =\frac{mU_1}{\left(m-n\right)U_1 +n}.
\end{equation}
As a result, a relation between $Y_1$ and $Y_2$ can be figured out from two equations, ${dW_1}/{d\alpha}={dW_2}/{d\alpha}=0$, as
\begin{equation} \label{dynamical-13}
Y_2 =\frac{\lambda \left(1-U_1\right)}{U_1}Y_1,
\end{equation}
along with an equation
\begin{equation}\label{dynamical-14}
3X^2 +\frac{1}{2} \left(Y_1^2 +\omega Y_2^2 \right)+\frac{1}{3} Z^2  =-\frac{\lambda}{2}Y_1.
\end{equation}
Here, $Z^2=Z_1^2+Z_2^2$ as an additional variable introduced for convenience. 
Additionally, another relation between $Y_1$ and $Y_2$ can be figured out from two other equations, ${dZ_1}/{d\alpha}={dZ_2}/{d\alpha}=0$, as
\begin{equation} \label{dynamical-15}
Y_2 =\frac{\rho \left(1-U_2\right)}{U_2}Y_1,
\end{equation}
along with a relation between $X$ and $Y_1$ defined as
\begin{equation}\label{dynamical-16}
2X+\left(\frac{\lambda}{2} +\rho \right) Y_1+2=0,
\end{equation}
with the help of Eq. \eqref{dynamical-14}. Here, it is noted that all $W_1$, $W_2$, $Z_1$, and $Z_2$ have been regarded as non-vanishing variables, similar to $U_1$ and $U_2$ as well as $Y_1$ and $Y_2$. Interestingly, three relations shown in Eqs. \eqref{dynamical-11}, \eqref{dynamical-13}, and \eqref{dynamical-15} imply that
\begin{align} \label{dynamical-17-1}
\frac{\rho}{\lambda}=\frac{m}{n} &=\kappa_1,\\
\label{dynamical-17-2}
n\rho =m\lambda &=\kappa_2,
\end{align}
which are nothing but that shown in Eqs. \eqref{relation-2-1} and \eqref{relation-2-2} in the previous section for the power-law solutions. Additionally, it appears from the equations, $dW_1/d\alpha=dW_2/d\alpha=0$, that
\begin{equation} \label{dynamical-18}
L= -\frac{n}{\lambda}= - \frac{m}{\rho},
\end{equation}
with the help of Eqs. \eqref{dynamical-13} and \eqref{dynamical-15}. This relation is identical to that shown in Eq. \eqref{L-relation} in the previous section for the power-law solutions. It is straightforward  to have from the relation \eqref{dynamical-11} that
\begin{equation} \label{dynamical-19}
\frac{U_2}{1-U_2} =\kappa_1 \frac{U_1}{1-U_1}.
\end{equation}
As a result, two equations, ${dY_1}/{d\alpha}={dY_2}/{d\alpha}=0$, imply an equation, 
\begin{equation} \label{dynamical-20}
\left(\frac{\lambda}{2}Y_1 +3 \right)\left\{ \left[ \frac{\omega}{\bar\lambda^2} n \lambda \rho +m   \right] Y_1 + \kappa_2 \right\}-   \left(\frac{\kappa_2}{6}+ m \rho  \right) Z^2 =0,
\end{equation}
with the help of Eqs. \eqref{dynamical-constraint}, \eqref{dynamical-13}, \eqref{dynamical-14}, \eqref{dynamical-17-1}, \eqref{dynamical-17-2}, and \eqref{dynamical-19}. It is noted that we have used the result that
\begin{equation}
\bar\lambda = \frac{U_1}{1-U_1}.
\end{equation}
Now, the equation $dX/d\alpha=0$, leads to an equation,
\begin{equation}\label{dynamical-21}
\left(\frac{\lambda}{2}Y_1 +3 \right) X -\frac{1}{3}Z^2 =0.
\end{equation}
For convenience, we will rewrite Eq. \eqref{dynamical-16} as 
\begin{equation} \label{dynamical-22}
2X+ \lambda \left(\kappa_1+ \frac{1}{2}   \right) Y_1+2=0.
\end{equation}
Up to now, we have derived three equations of three variables $X$, $Y_1$, and $Z$, Eqs. \eqref{dynamical-20}, \eqref{dynamical-21}, and \eqref{dynamical-22}. As a result, solving these equations gives us a non-trivial solution,
\begin{align}\label{dynamical-23}
X&=\frac{2\left[ \kappa_2 \lambda \bar\lambda^2  \left(2\kappa_1 +1 \right)-4\left(\omega n \lambda \rho +m \bar\lambda^2 \right) \right]}{\lambda \bar\lambda^2 \left(2\kappa_1 +1 \right) \left(\kappa_2 +6m\rho \right)+8\left(\omega n\lambda \rho +m\bar\lambda^2 \right)}, \\
\label{dynamical-24}
Y_1&=\frac{-12 \bar\lambda^2 \left(\kappa_2 +2m\rho \right)}{\lambda \bar\lambda^2 \left(2\kappa_1 +1 \right) \left(\kappa_2 +6m\rho \right)+8\left(\omega n\lambda \rho +m\bar\lambda^2 \right)},\\
\label{dynamical-25}
Z^2 &= \frac{18 \left[ \kappa_2 \lambda \bar\lambda^2  \left(2\kappa_1 +1 \right)-4\left(\omega n\lambda \rho +m \bar\lambda^2 \right)\right] \left\{ \lambda \bar\lambda^2 \left[ 2m\rho \left(6\kappa_1 +1 \right) +\kappa_2 \left(2\kappa_1 -1 \right)\right] +8 \left(\omega n \lambda \rho +m\bar\lambda^2 \right)\right\}}{\left[\lambda \bar\lambda^2 \left(2\kappa_1 +1 \right) \left(\kappa_2 +6m\rho \right)+8\left(\omega n\lambda \rho +m\bar\lambda^2 \right) \right]^2},
\end{align}
 where we have ignored the isotropic fixed points corresponding to $X=0$. It is noted that two types of isotropic fixed points, which have been found in Ref. \cite{Chen:2021nkf}, can be easily derived from the above dynamical system. Indeed,  by setting $Y_1 \neq 0$, $W_1 \neq 0$, and $Y_2=W_2=U_1=U_2=Z_1=Z_2=0$ we can obtain the corresponding isotropic slow-roll inflation; while $Y_1 \neq 0$, $W_1 \neq 0$, and $W_2=U_1=U_2=Z_1=Z_2=0$ but $Y_2 \neq 0$ will lead to the corresponding  isotropic hyperbolic inflation. Interestingly, one more isotropic fixed point can also be figured out in the present model, which corresponds to $U_2=Z_1=Z_2=0$ along with  $Y_1 \neq 0$, $Y_2 \neq 0$, $W_1\neq 0$, $W_2 \neq 0$, and $U_1 \neq 0$. In fact, it is a generalised isotropic hyperbolic   inflation with
\begin{align}
Y_1 =-\frac{\lambda \bar\lambda^2}{\omega \lambda^2 + \bar\lambda^2};~Y_2 = \frac{\lambda}{\bar\lambda}Y_1.
\end{align}
In addition to the above anisotropic fixed point, it should be noted that another anisotropic fixed point with $X \neq 0$, which is nothing but the anisotropic slow-roll inflation \cite{MW,Chen:2021nkf}, can be derived in this paper by setting $Y_2=W_2=U_1=U_2=Z_2=0$. However, it will not be our current interest because of the fact that it is not equivalent to the anisotropic power-law solution found in the previous section.  One can now ask if the anisotropic fixed point shown in Eqs. \eqref{dynamical-23}, \eqref{dynamical-24}, and \eqref{dynamical-25} is equivalent to the anisotropic power-law found in the previous section. In order to answer this question, we will rewrite $X$, $Y_1$, and $Z^2$ as
 \begin{align}
 \label{dynamical-26}
 X&= \frac{2\left[ \kappa_2 n \lambda \left(2\kappa_1 +1 \right) -4 \left(\omega u_0 \lambda\rho  +mn \right)\right]}{n\lambda \left(2\kappa_1 +1 \right) \left(\kappa_2 +6m\rho \right) +8 \left(\omega u_0 \lambda \rho +mn \right)},\\
  \label{dynamical-27}
 Y_1&= \frac{-12n \left(\kappa_2 +2m\rho \right)}{n\lambda \left(2\kappa_1 +1 \right) \left(\kappa_2 +6m\rho \right) +8 \left(\omega u_0 \lambda \rho +mn \right)},\\
  \label{dynamical-28}
  Z^2&=\frac{18 \left[\kappa_2 n\lambda \left(2\kappa_1+1\right)- 4 \left(\omega u_0 \lambda \rho + mn \right) \right] \left\{ n\lambda \left[ 2m\rho \left(6\kappa_1 +1 \right) +\kappa_2 \left(2\kappa_1 -1 \right)\right] +8 \left(\omega u_0 \lambda \rho +mn \right) \right\} }{\left[ n\lambda \left(2\kappa_1 +1 \right) \left(\kappa_2 +6m\rho \right) +8 \left(\omega u_0 \lambda \rho +mn \right)\right]^2 },
 \end{align}
with the help of useful relations,
  \begin{equation} \label{dynamical-26}
 \bar\lambda=\frac{ n}{\sqrt{u_0}};~\bar\rho =\frac{ m}{\sqrt{u_0}},
 \end{equation}
 here $u_0$ has been defined in Eq. \eqref{definition-of-u0}. Now, it is clear that this  anisotropic fixed point is absolutely equivalent to the anisotropic power-law solution found above. Indeed, one can easily check that $X = \eta /\zeta$ with $\zeta$ and $\eta$ have been shown in Eqs. \eqref{solution of zeta} and \eqref{solution of eta}, respectively. As a result, the anisotropic fixed point can be approximated during the inflationary phase with $\rho \gg \lambda  $, $|m|\gg |n|$, and $\kappa_1 \gg 1$ as
 \begin{align}
 X & \simeq \frac{1}{3\kappa_1} \ll 1;~Y_1 \simeq -\frac{2}{\rho};~Y_2 \simeq -\frac{2}{m} \sqrt{u_0},\\
 Z^2 &\simeq 9X \ll 1;~W_1^2 +W_2^2 \simeq 3,\\
 U_1 &= \frac{n}{n+\sqrt{u_0}};~U_2 = \frac{m}{m+\sqrt{u_0}},
 \end{align}
 here we have assumed that $u_0 \sim {\cal O}(1)$. It appears that $Z^2 \ll 1$ implies that $Z_1 \ll 1$ along with $Z_2 \ll 1$. Additionally, the result $W_1^2+W_2^2 \simeq 3$ indicates that $0 <W_1, ~W_2 <\sqrt{3}$.
 Next, we will investigate the stability of the obtained anisotropic fixed point, similar to our previous paper \cite{Do:2021lyf}. In particular, we will perturb the dynamical system around the fixed point as follows
\begin{align}
\frac{d \delta X}{d\alpha} \simeq & -3\delta X,\\
\frac{d \delta Y_1}{d\alpha} \simeq & -3\delta Y_1 +  \frac{2\omega}{L}Y_2 \delta Y_2 +2\rho Z_1 \delta Z_1 -2\lambda W_1 \delta W_1 ,\\
\frac{d \delta Y_2}{d\alpha} \simeq & - \frac{Y_2}{L}\delta Y_1 - \left(\frac{Y_1}{L}+3 \right) \delta Y_2 +\frac{1}{\omega} \left[ \frac{2Z_2 U_2}{1-U_2}  \delta Z_2 +\left(\frac{Z_2}{1-U_2}\right)^2 \delta U_2 \right]\nonumber\\
&-\frac{1}{\omega} \left[ \frac{2 W_2 U_1}{1-U_1}  \delta W_2 +\left(\frac{W_2}{1-U_1}\right)^2 \delta U_1 \right],\\
\frac{d\delta Z_1}{d\alpha} \simeq& -Z_1 \left( 2\delta X +\rho \delta Y_1 \right),\\
\frac{d\delta Z_2}{d\alpha}\simeq& -Z_2 \left[ 2\delta X +\frac{U_2}{1-U_2} \delta Y_2 +\frac{Y_2}{\left(1-U_2 \right)^2} \delta U_2 \right],\\
\frac{d\delta W_1}{d\alpha} \simeq&~\frac{\lambda}{2}W_1 \delta Y_1,\\
\frac{d\delta W_2}{d\alpha} \simeq&~W_2 \left[ \frac{U_1}{2 \left(1-U_1 \right)} \delta Y_2 +\frac{Y_2}{2\left(1-U_1\right)^2} \delta U_1 \right],
\end{align}
\begin{align}
\frac{d\delta U_1}{d\alpha}\simeq&-U_1^2 \left(\frac{1-U_1}{U_1}\frac{\delta Y_1}{L}+\frac{\delta Y_2}{n}-\frac{Y_1}{U_1^2}\frac{\delta U_1}{L} \right),\\
\frac{d\delta U_2}{d\alpha}\simeq&-U_2^2 \left(\frac{1-U_2}{U_2}\frac{\delta Y_1}{L}+\frac{\delta Y_2}{m} -\frac{Y_1}{U_2^2}\frac{\delta U_2}{L} \right).
 \end{align}
 Taking exponential perturbations \cite{Do:2021lyf},
 \begin{align}
&  \delta X = A_1 \exp[\tau \alpha]; ~ \delta Y_1 = A_2 \exp[\tau \alpha];~ \delta Y_2 = A_3 \exp[\tau \alpha],\\
&  \delta Z_1 = A_4 \exp[\tau \alpha];~ \delta Z_2 = A_5 \exp[\tau \alpha]; ~ \delta W_1 = A_6 \exp[\tau \alpha],\\
& \delta W_2 = A_7 \exp[\tau \alpha]; ~\delta U_1 =A_8 \exp[\tau\alpha];~\delta U_2 =A_9 \exp[\tau\alpha],
 \end{align}
we are able to write the above perturbed equations as a homogeneous linear system of $A_i$ with $i=1-9$. Furthermore, this system can be written as a homogeneous matrix equation as follows
\begin{equation} \label{stability-equation}
{\cal M}\left( {\begin{array}{*{20}c}
   A_{1}  \\
   A_{2}  \\
   A_{3} \\
   A_{4}  \\
   A_{5}  \\
   A_{6}\\
   A_{7}\\
   A_{8}\\
   A_{9}\\
 \end{array} } \right) =0,
 \end{equation}
 where the matrix ${\cal M}$ is given by
 \begin{equation}
 {\cal M} \equiv \left[ {\begin{array}{*{20}c}
   {-3-\tau} & {0} & {0 } & {0 } & {0} &{0} &{0}&{0}&{0} \\
   {0 } & {-3-\tau} & { \frac{2\omega}{L}Y_2 } & {2\rho Z_1} &{0}&{-2\lambda W_1}&{0}&{0}&{0}  \\
     {0 } & {- \frac{Y_2}{L}} & {- \left(\frac{Y_1}{L}+3 \right)-\tau } & {0 } &{\frac{2Z_2 U_2}{\omega \left(1-U_2 \right)}  }&{0}&{\frac{2W_2 U_1}{\omega \left(U_1-1 \right)}} &{\frac{-1}{\omega} \left(\frac{W_2}{1-U_1}\right)^2  }&{\frac{1}{\omega} \left(\frac{Z_2}{1-U_2} \right)^2} \\
   {-2Z_1} & {-\rho Z_1 } & {0 } & {-\tau } &{0}&{0}&{0} &{0}&{0} \\
   {-2Z_2} & {0 } & {\frac{Z_2 U_2}{U_2-1}  } &{0}& {-\tau } &{0}&{0} &{0}&{\frac{-Y_2 Z_2}{\left(1-U_2 \right)^2}} \\
   {0}&{\frac{\lambda }{2}W_1}&{0} &{0} &{0}&{-\tau}&{0} &{0}&{0} \\
{0}&{0}&{\frac{W_2 U_1}{2\left(1-U_1\right)}} &{0} &{0}&{0}&{-\tau} &{\frac{Y_2 W_2}{2 \left(1-U_1 \right)^2}}&{0} \\
{0}&{\frac{U_1 \left(U_1-1\right)}{L}}&{\frac{-U_1^2}{n}} &{0} &{0}&{0}&{0} &{\frac{Y_1}{L}-\tau}&{0} \\
{0}&{\frac{U_2 \left(U_2-1 \right)}{L}}&{\frac{-U_2^2}{m}} &{0} &{0}&{0}&{0} &{0}&{\frac{Y_1}{L}-\tau} \\
 \end{array} } \right].
\end{equation}
Mathematically, this homogeneous linear system will admit non-trivial solutions, i.e., there is at least one non-vanishing solution $A_i \neq 0$, if and only if 
\begin{equation}
\det {\cal M} =0.
\end{equation}
As a result, this equation can be written as an equation of $\tau$ as
\begin{equation} \label{equation-of-m}
   \tau^3 \left(\tau+3 \right) \left(m\tau-2 \right) \left(a_4 \tau^4 +a_3 \tau^3 +a_2 \tau^2 +a_1 \tau +a_0 \right)=0,
\end{equation}   
where the coefficients, $a_i$ ($i=0-4$), are given by
\begin{align}
a_4&= \omega u_0 m^5 ,\\
a_3&=6\omega u_0 m^5,\\
a_2 &\simeq 2 m^5 \left(m^2 Z_2^2 + \omega u_0 \rho^2 Z_1^2  \right) ,\\
a_1&\simeq 6m^5 \left(m^2 Z_2^2 + \omega u_0 \rho^2 Z_1^2 \right)  ,\\
a_0 &\simeq  2 m^7 \rho^2 Z_1^2 Z_2^2 <0,
\end{align}
thanks to the approximations of the anisotropic fixed points.
Here, we have only kept the leading terms of the coefficients $a_i$ ($i=0-2$) due to a set of the corresponding constraints for the anisotropic inflationary solution, $\rho \gg \lambda $ and $|m| \gg |n|$ as well as $\lambda>0, ~\rho >0$ and $n<0, ~m <0$, for simplicity. It turns out that besides five non-positive roots, $\tau_{1,2,3}=0$, $\tau_4=-3<0$, and $\tau_5 =2/m <0$, four other non-trivial roots of Eq. \eqref{equation-of-m} are derived from the equation
\begin{equation} \label{equation-of-m-2}
{\cal F} (\tau)\equiv a_4 \tau^4 +a_3 \tau^3 +a_2 \tau^2 +a_1 \tau +a_0 =0.
\end{equation}
 It turns out that if $\omega =+1 >0$ then all coefficients $a_i$ ($i=0-4$) become negative provided that $u_0>0$. Mathematically, Eq. \eqref{equation-of-m-2} with all negative coefficients $a_i$ ($i=0-4$) no longer admits any positive roots $\tau >0$, meaning that the corresponding anisotropic power-law hyperbolic inflationary solution is indeed stable against field perturbations. However, if $\omega =-1 <0$, i.e., $\psi$ is the phantom-like scalar field, then $a_4 >0$ while $a_0 <0$. Consequently, Eq. \eqref{equation-of-m-2} will admit at least one positive root $\tau >0$. This fact can be easily verified by an observation that the curve ${\cal F}(\tau)$ with ${\cal F}(0)=a_4 <0$ and ${\cal F} (\tau \gg 1) \sim a_4\tau^4 >0$ will cross the positive horizontal $\tau$-axis  at least one time at $\tau=\tau_\ast >0 $. The existence of positive $\tau_\ast$ implies that the corresponding anisotropic power-law hyperbolic inflationary solution is indeed unstable.  This result is consistent with our previous models \cite{WFK,Do:2017rva,Do:2021lyf}, in which the phantom field has been shown to favor the cosmic no-hair conjecture by  causing unstable mode(s) to the corresponding anisotropic power-law inflationary solutions. 
 
\begin{figure}[hbtp] 
\begin{center}
{\includegraphics[height=85mm]{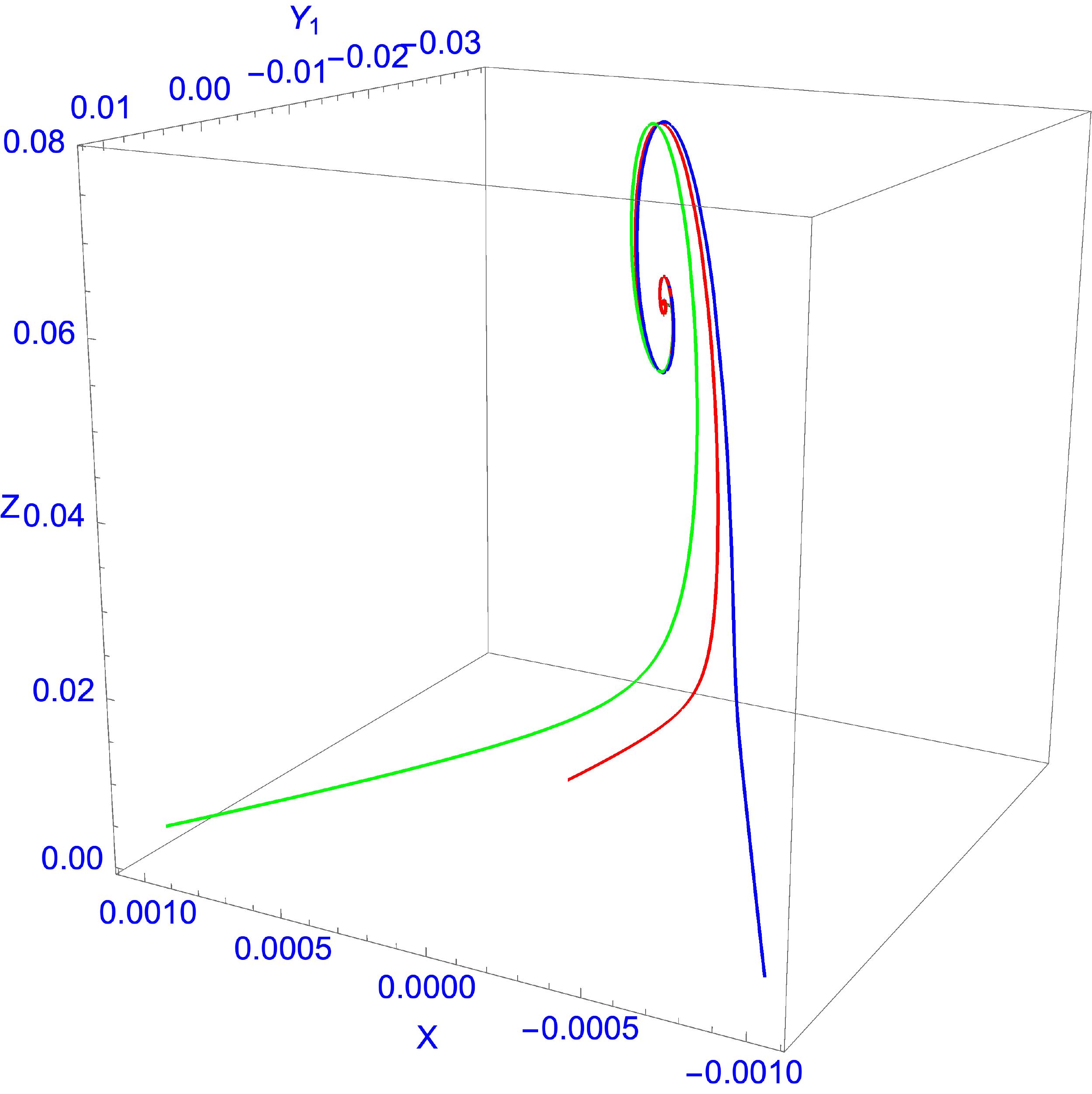}}\quad
{\includegraphics[height=85mm]{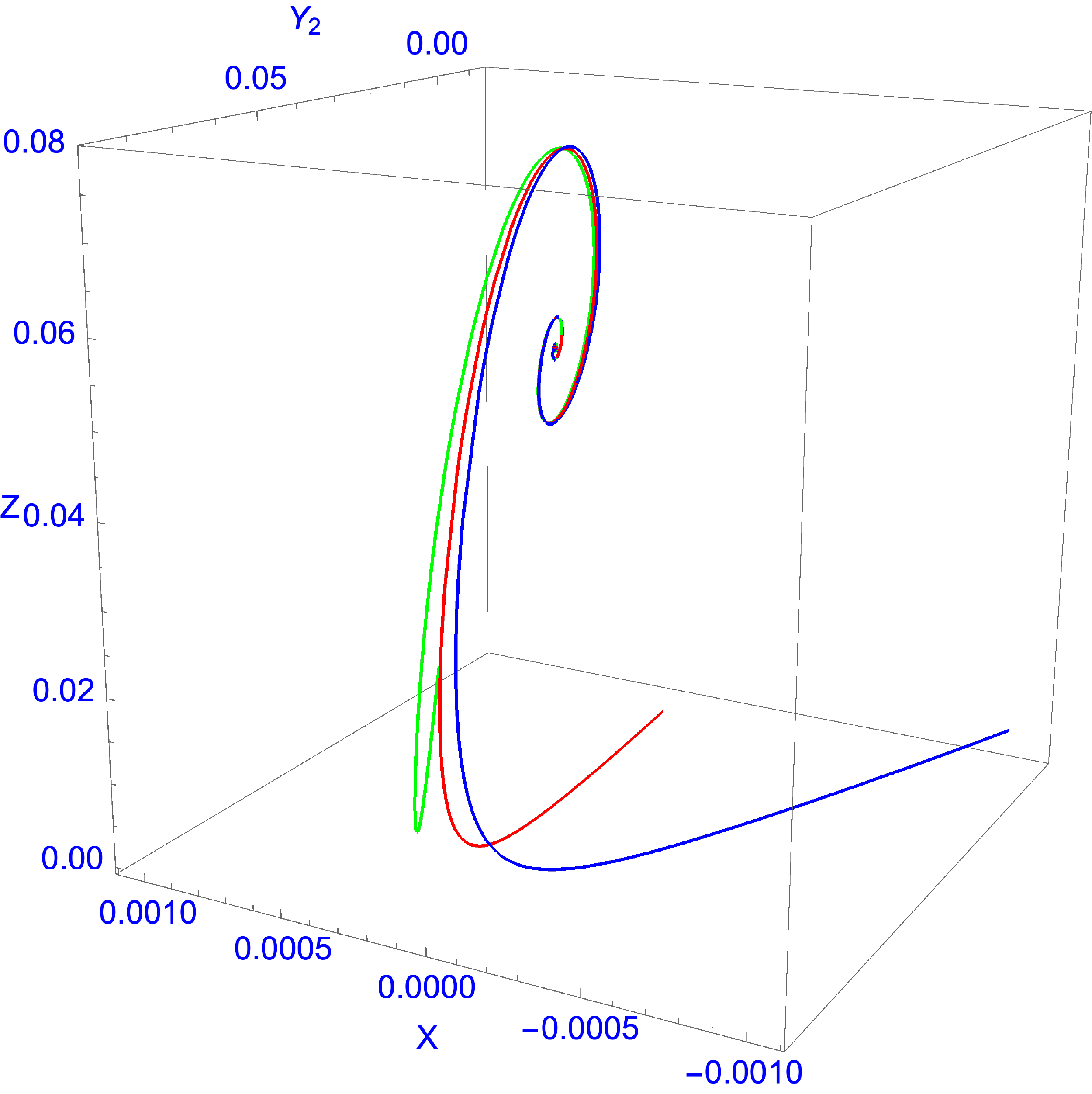}}\\
\caption{Attractor behavior of the anisotropic fixed point with $\omega=+1$, $u_0=2$, $\lambda=0.1$, $\rho =50$, $n=-0.2$, and $m=-100$. These two plots clearly show that three different trajectories corresponding to different initial conditions all tend to converge to the anisotropic fixed point.}
\label{fig1}
\end{center}
\end{figure}

It should be noted that, we are able to numerically show the attractor property of the anisotropic fixed point with $\omega=+1$ (see Fig. \ref{fig1} for details). This result acts as a strong confirmation of the stability of the anisotropic power-law hyperbolic inflationary solution. It should be noted that if the angular field $\psi$ is the phantom-like scalar field, i.e., $\omega =-1$, the corresponding anisotropic fixed point will be unattractive as expected since all trajectories tend to converge to the isotropic fixed point corresponding to $X=Z=0$. These results are also consistent with our previous study \cite{Do:2021lyf}.
\section{Conclusions} \label{final}
Motivated by a recent paper \cite{Chen:2021nkf}, we have proposed a hyperbolic generalization of our recent model, which acts as a novel multifield extension of the KSW model \cite{Do:2021lyf}. In this generalization, the field space of two scalar fields are assumed to be a hyperbolic space instead of a conventional flat space \cite{Brown:2017osf}. One of the scalar fields, $\phi$, is the radial field and the other, $\psi$, is the angular field. Both of them are massive and coupled to vector fields. As a result, we have obtained a set of Bianchi type I power-law solutions to this model in the regime $\phi \gg L$, similar to Ref.  \cite{Chen:2021nkf}. Furthermore, we have shown that this solution can be used to present an anisotropic inflationary solution if $\rho \gg \lambda$ along with $|m| \gg |n|$, provided that both $\lambda$ and $\rho$ are positive, while both $n$ and $m$ are negative. Stability analysis based on the dynamical system method has been performed to verify that this anisotropic inflationary solution is indeed stable and attractive, similar to the solutions obtained in the non-hyperbolic (flat) model \cite{Do:2021lyf}. However, if the angular field, $\psi$, is the phantom-like one with $\omega=-1$, the corresponding anisotropic inflationary solution will be unstable as expected. It should be noted that this present paper is solely devoted to seek anisotropic inflationary solutions and investigate their stability in order to deal with the cosmic no-hair conjecture. Detailed investigations on the CMB imprints \cite{Imprint1} of this model  will be presented elsewhere. We hope that our present paper would be useful to studies of the early time universe.
\begin{acknowledgments}
We would like to thank the referee very much for useful comments and suggestions. T.Q.D. is supported by the Vietnam National Foundation for Science and Technology Development (NAFOSTED) under grant number 103.01-2020.15. W.F.K. is supported in part by the Ministry of Science and Technology (MOST) of Taiwan under Contract No. MOST 110-2112-M-A49-007. 
\end{acknowledgments}

\end{document}